\PassOptionsToPackage{breaklinks}{hyperref}
\documentclass[sigconf,screen,balance,natbib=false,timestamp=false,urlbreakonhyphens=true]{acmart}

\usepackage[T1]{fontenc}
\usepackage[utf8]{inputenc}
\usepackage[capitalize,noabbrev]{cleveref} %
\usepackage{microtype}
\usepackage{graphicx}
\usepackage{tabularx}
\usepackage{booktabs}
\usepackage[shortlabels,inline]{enumitem}
\usepackage{xspace}
\usepackage{balance}
\usepackage[colorinlistoftodos,prependcaption,textsize=scriptsize]{todonotes} %
\usepackage[type={CC}, modifier={by}, version={4.0}]{doclicense}

\usepackage{listings}
\usepackage[abbreviate=true, dateabbrev=true, alldates=year, isbn=false, doi=false, urldate=comp, url=true, maxbibnames=9, backref=false, backend=biber, style=numeric-comp, language=american, giveninits=true, sortcites=true, sorting=none]{biblatex}
\AtEveryBibitem{\clearlist{location}} %
\AtEveryBibitem{\clearfield{series}} %
\AtEveryBibitem{\clearlist{language}} %

\DeclareSourcemap{ 
  \maps[datatype=bibtex]{
    \map{
      \pertype{article}
      \pertype{inproceedings}
      \pertype{incollection}
      \pertype{book}
      \pertype{techreport}
      \pertype{phdthesis}
      \step[fieldset=url, null]
      \step[fieldset=urldate, null]
    }
  }
}

\DeclareSourcemap{
  \maps[datatype=bibtex]{
    \map{
      \step[fieldsource=booktitle, match={\{\{}, replace={}] 			
      \step[fieldsource=booktitle, match={\}\}}, replace={}] 	
      \step[fieldsource=booktitle, match=\regexp{^Proceedings\s+of\s+the\s+\S+\s+}, replace={}]
      \step[fieldsource=booktitle, match=\regexp{^\d\d\d\d\s+}, replace={}]
      \step[fieldsource=booktitle, match=\regexp{^[\d]{1,2}th\s+}, replace={}]
      \step[fieldsource=booktitle, match=\regexp{^[\d]{1,2}nd\s+}, replace={}]
      \step[fieldsource=booktitle, match={International}, replace={Int'l}]
      \step[fieldsource=booktitle, match=\regexp{[\{]{0,2}Conference[\}]{0,2}\s+on(\s+the)?\s+}, replace=\regexp{Conf.\\\x{20}}]
      \step[fieldsource=booktitle, match={Conference }, replace={Conf.\ }]
      \step[fieldsource=booktitle, match=\regexp{[\{]{0,2}Symposium[\}]{0,2}\s+on(\s+the)?\s+}, replace=\regexp{Symp.\\\x{20}}]
      \step[fieldsource=booktitle, match={Symposium }, replace={Symp.\ }]
      \step[fieldsource=booktitle, match=\regexp{[\{]{0,2}Workshop[s]?[\}]{0,2}\s+on(\s+the)?\s+}, replace=\regexp{Ws.\\\x{20}}]
      \step[fieldsource=booktitle, match=\regexp{Workshop[s]?}, replace={Ws.\ }]
      \step[fieldsource=booktitle, match=\regexp{[\{]{0,2}Foundations[\}]{0,2}\s+of(\s+the)?\s+}, replace=\regexp{Found.\\\x{20}}]
      \step[fieldsource=booktitle, match={Automated }, replace={Autom.\ }]
      \step[fieldsource=booktitle, match={Software }, replace={Softw.\ }]
      \step[fieldsource=booktitle, match={Requirements }, replace={Req.\ }]
      \step[fieldsource=booktitle, match={Engineering}, replace={Eng.}]
      \step[fieldsource=booktitle, match={Applications}, replace={Applic.}]
      \step[fieldsource=booktitle, match=\regexp{(.*)\s+\(.*\)$}, replace=\regexp{$1}]
      \step[fieldsource=journal, match=\regexp{(The\s+)?Journal\s+of\s+}, replace=\regexp{J.\\\x{20}}]
      \step[fieldsource=journal, match=\regexp{Communications\s+of\s+the\s+}, replace=\regexp{Comm.\\\x{20}}]
      \step[fieldsource=journal, match=\regexp{(ACM|IEEE)\s+Transactions\s+on\s+(.*)}, replace=\regexp{$1\x{20}Trans.\\\x{20}$2}]
      \step[fieldsource=journal, match={Software }, replace={Softw.\ }]
      \step[fieldsource=journal, match={Engineering}, replace={Eng.}]
      \step[fieldsource=journal, match={Applications}, replace={Applic.}]
      \step[fieldsource=journal, match=\regexp{(.*)\s+\(.*\)$}, replace=\regexp{$1}]
      \step[typesource=techreport, typetarget=report]
    }
  }
}

\usepackage{pgfplotstable}

\newcommand{\CVEfixes}{\emph{CVEfixes}\xspace}
\newcommand{\head}[1]{\par\noindent\textbf{#1}}
\newcommand{\ERtable}[1]{\emph{\textsf{#1}}}

\newcommand{\jsonPullDate}{9 June 2021\xspace} %
\newcommand{\cveCount}{5365\xspace}		%
\newcommand{\cweCount}{180\xspace}	%
\newcommand{\projectCount}{1754\xspace}	%
\newcommand{\commitCount}{5495\xspace}			%

\newcommand{\commitFirstCount}{1029\xspace}%

\newcommand{\cveFirst}{\emph{linux}\xspace}
\newcommand{\cveSecond}{\emph{ImageMagick}\xspace}

\newcommand{\cveFirstCount}{973\xspace}%
\newcommand{\cveSecondCount}{157\xspace}%

\newcommand{\percentCveTen}{31.84\%\xspace}	    %
\newcommand{\percentCommitTen}{31.61\%\xspace}	%
\newcommand{\percentFileTen}{42.43\%\xspace}		%
\newcommand{\percentMethodTen}{57.84\%\xspace}	%

\newcommand{\percentCveCweTen}{55.58\%\xspace}	  %
\newcommand{\percentCommitCweTen}{57\%\xspace}	  %
\newcommand{\percentFileCweTen}{55.75\%\xspace}	%

\newcommand{\LM}[2][]{\todo[linecolor=orange,backgroundcolor=orange!40,bordercolor=orange,#1]{#2}} %
  {\newsavebox{\myStore}\LM{#1}\begin{lrbox}{\myStore}\hspace*{-1ex}\begin{minipage}{\textwidth-1ex}}%
  {\end{minipage}\end{lrbox}\LM[inline,nolist]{\usebox{\myStore}}}
\newcommand{\GB}[2][]{\todo[linecolor=green,backgroundcolor=green!25,bordercolor=green,#1]{#2}}

  {\newsavebox{\myStore}\NP{#1}\begin{lrbox}{\myStore}\hspace*{-1ex}\begin{minipage}{\textwidth-1ex}}%
  {\end{minipage}\end{lrbox}\GB[inline,nolist]{\usebox{\myStore}}}
\newcommand{\AG}[2][]{\todo[linecolor=blue,backgroundcolor=blue!25,bordercolor=blue,#1]{#2}}

  {\newsavebox{\myStore}\NP{#1}\begin{lrbox}{\myStore}\hspace*{-1ex}\begin{minipage}{\textwidth-1ex}}%
  {\end{minipage}\end{lrbox}\AG[inline,no list]{\usebox{\myStore}}}

\pgfplotsset{compat=1.17}
\pgfplotstableset{
    col sep = comma,
    header=true,
    every head row/.style={before row=\toprule,after row=\midrule},
    every last row/.style={after row=\bottomrule},
    assign column name/.style={/pgfplots/table/column name={\textbf{#1}}},
}
\newcommand{\tablefont}{\small} %

\setcopyright{rightsretained}
\acmPrice{}
\acmDOI{10.1145/3475960.3475985}
\acmYear{2021}
\copyrightyear{2021}
\acmSubmissionID{fsews21promisemain-p11-p}
\acmISBN{978-1-4503-8680-7/21/08}
\acmConference[PROMISE '21]{Proceedings of the 17th International Conference on Predictive Models and Data Analytics in Software Engineering}{August 19--20, 2021}{Athens, Greece}
\acmBooktitle{Proceedings of the 17th International Conference on Predictive Models and Data Analytics in Software Engineering (PROMISE '21), August 19--20, 2021, Athens, Greece}

\makeatletter
\def\@copyrightpermission{\doclicenseImage[imagewidth=2cm]\hspace*{1mm}\raisebox{3mm}{\parbox{6.3cm}{This work is licensed under a \doclicenseLongNameRef{} 
(\doclicenseNameRef) license.}}\vspace{1ex}}%
\makeatother

\begin{document}

\title{\CVEfixes: Automated Collection of Vulnerabilities and Their Fixes from Open-Source Software}

\author{Guru Bhandari}
\affiliation{%
  \institution{Simula Research Laboratory}
  \city{Oslo}
  \country{Norway}
}
\email{guru@simula.no}

\author{Amara Naseer}
\affiliation{%
  \institution{Simula Research Laboratory}
  \city{Oslo}
  \country{Norway}
}
\email{amara@simula.no}

\author{Leon Moonen}
\affiliation{%
  \institution{Simula Research Laboratory}
  \city{Oslo}
  \country{Norway}
}
\email{leon.moonen@computer.org}

\makeatletter
\if@ACM@anonymous
  \gdef\addresses{\@author{\vspace*{0.35em}Anonymous Author(s)\vspace*{2em}}}
\fi
\makeatother


\noindent

\begin{abstract}

Data-driven research on the automated discovery and repair of security vulnerabilities in source code 
requires comprehensive datasets of real-life vulnerable code and their fixes.
To assist in such research, 
we propose a method to automatically collect and curate a comprehensive vulnerability dataset from 
Common Vulnerabilities and Exposures (CVE) records in the public National Vulnerability Database (NVD).
We implement our approach in a fully automated dataset collection tool
and share an initial release of the resulting vulnerability dataset named \CVEfixes.

The \CVEfixes collection tool automatically fetches all available CVE records from the NVD, 
gathers the vulnerable code and corresponding fixes from associated open-source repositories,
and organizes the collected information in a relational database.
Moreover, 
the dataset is enriched with meta-data such as programming language, 
and detailed code and security metrics at five levels of abstraction.
The collection can easily be repeated to keep up-to-date with newly discovered or patched vulnerabilities. 
The initial release of \CVEfixes spans all published CVEs up to \jsonPullDate, 
covering \cveCount CVE records for \projectCount open-source projects
that were addressed in a total of \commitCount vulnerability fixing commits.

\CVEfixes supports various types of data-driven software security research,
such as vulnerability prediction, 
vulnerability classification,
vulnerability severity prediction,
analysis of vulnerability-related code changes,
and automated vulnerability repair.

\end{abstract}

 
\maketitle

\section{Introduction}

\noindent
The exploitation of security vulnerabilities is a significant threat to the reliability of software systems
and to the protection of the data processed by them, 
as is frequently evidenced by new data leaks, ransomware attacks, and substantial outages of essential systems.
Despite the continued efforts of the software engineering community to improve software quality and security 
by means of secure coding guidelines, software testing, and various forms of code review,
the publicly available Common Vulnerabilities and Exposures (CVE) records 
reveal an increasing trend in the number of vulnerabilities that are discovered each year~\cite{mitre:cve}.

The overall security of software is highly dependent on the effective and timely identification
and mitigation of software vulnerabilities.
However, this is not an easy task and requires experience 
and specialized skills that go beyond the expertise of the typical developer,
resulting in many vulnerabilities that go unnoticed for a long time.
Consequently, there has been considerable attention in academia and industry 
to the development of techniques and tools that can help developers identify, 
and possibly repair, security vulnerabilities in source code already in the development phase.

\emph{Data-driven} vulnerability research depends on the availability of datasets
with samples of \emph{real-life} vulnerable code and their fixes~\cite{choi2017:endtoend}.
Moreover, %
such datasets should encompass multiple levels of \emph{granularity}~\cite{morrison2015:challenges,zou2019:mvuldeepecker}, 
such as files, classes, functions, etc.,
and cover widely-used programming languages~\cite{wang2019:detecting,feng2020:codebert}.
Finally, for reliable training and evaluation of machine learning (ML) approaches, 
we need \emph{comprehensive} datasets that contain large numbers of diverse 
and \emph{labeled} samples of both vulnerable and non-vulnerable code~\cite{russell2018:automated,coulter2020:code,lin2020:software}.
As we will see in the discussion of related work (\cref{sec:relwork}), 
the currently available vulnerability datasets do not fulfill these requirements.

To address their shortcomings and assist in data-driven vulnerability research, 
we propose to automatically collect and curate a comprehensive 
vulnerability dataset from CVE records in the National Vulnerability Database (NVD). 
In particular, 
we propose an approach that \emph{mines} CVE records for open-source software (OSS) projects hosted on GitHub, GitLab, and Bitbucket, 
to collect real-world samples of vulnerable and corresponding patched code.
We implement the proposed approach in a dataset collection tool
and share an initial release of the resulting vulnerability dataset named \CVEfixes.
The dataset supports various types of data-driven software security research, 
such as ML-based vulnerability identification, 
automated classification of identified vulnerabilities in Common Weakness Enumeration (CWE) types,
vulnerability severity prediction, 
and automated repair of security vulnerabilities.

\head{Contributions:} This paper makes the following contributions:
\begin{sloppypar}
\begin{enumerate}[(a), nosep]
\item We survey existing security vulnerability-related datasets and discuss their strengths and weaknesses.
\item We propose a method to automatically collect and curate a comprehensive vulnerability dataset from CVE records in the NVD, 
and obtain real-world samples of vulnerable and corresponding patched code, 
organized at multiple levels of granularity.
In addition, we discuss how to enrich this data with meta-data such as programming language used, 
and detailed code-related metrics at five levels of abstraction, such as the commit-, file-, and method levels,
as well as the repository- and CVE levels. 
\item We implement the proposed approach in a dataset collection tool that is made publicly available.\footnote%
{~https://github.com/secureIT-project/CVEfixes, \textsc{doi:}\href{https://doi.org/10.5281/zenodo.5111494}{10.5281/zenodo.5111494}.}
\item We publicly share a version of the \CVEfixes vulnerability dataset for use by other 
researchers.%
\footnote{~\url{https://zenodo.org/record/4476563}, \textsc{doi:}\href{https://doi.org/10.5281/zenodo.4476563}{10.5281/zenodo.4476563}.}
The initial release spans all published CVE records up to \jsonPullDate, 
covering \cveCount CVE records for \projectCount OSS projects. 
A total of \commitCount vulnerability fixing commits are obtained from the projects’ version control systems
and linked to information from the corresponding CVE records,
such as CVE-IDs, reference links, severity scores, vulnerability type/CWE type, and other descriptive information.
\end{enumerate}
\end{sloppypar}\noindent
The remainder of the paper is organized as follows: 
\cref{sec:relwork} discusses the currently available vulnerability datasets and how our work differs from them.
\cref{sec:construction} presents the process for constructing the \CVEfixes dataset,
followed by an exploration of the initial release in \cref{sec:statistics}.
\cref{sec:using} discusses applications of the dataset, 
together with current limitations and future extensions. 
Finally, we conclude in \cref{sec:concl}.

\section{Related Work}
\label{sec:relwork}

\noindent
Several vulnerability-related test suites and datasets were developed in the last decade, often with particular goals in mind. 
These sets were consequently picked up by other security researchers looking for relevant data to train or evaluate their techniques on.
This section surveys frequently used datasets and discusses some of the challenges for their reuse in other contexts. 
We also look at more general software repositories and repository mining frameworks.
Finally, we discuss how the \CVEfixes dataset is different from the existing work in this area.

\head{Vulnerability Datasets}
The Software Assurance Metrics And Tool Evaluation (SAMATE) project created the Juliet Test Suite 
as a benchmark for static analysis tools that aim to identify vulnerabilities in source code~\cite{boland2012:juliet,black2018:software}.
Programs from the Juliet Test Suite have been used in several studies on vulnerability
and weakness prediction~\cite{gupta2018:textmining, gupta2015:predicting, mokhov2015:marfcat, medeiros2014:automatic, mokhov2014:use}.
However, the benchmark was never created for this purpose, 
and concerns have been raised that the vulnerabilities in Juliet are not very diverse, 
and many of them are of a synthetic nature that does not occur in real-world software projects~\cite{choi2017:endtoend, zheng2021:d2a}.

Mitropoulus et al.~\cite{mitropoulos2014:vulnerability} constructed a vulnerability dataset by analyzing the Maven repository using
\emph{FindBugs}~\cite{:findbugs}, a tool conducting static code analysis on Java byte-code,
identifying the vulnerabilities in Maven and categorizing them into nine different types.
Similarly, Draper's VDISC dataset~\cite{russell2018:automated} consists of C/C++ source code of 1.27 million functions
mined from open-source projects that were labeled by three static code analysis tools (Clang, Cppcheck, and Flawfinder) as vulnerable or not,
classified in five groups of vulnerabilities, namely CWE-120, 119, 469, 476, and other.
However, there are some challenges with the dataset, 
such as the fact that it is highly imbalanced with only 6.8\% functions labeled as vulnerable.
Moreover, the extracted functions are incomplete, 
missing the function's return type which excludes certain (signature-based) analyses.
\emph{SVCP4C} (\emph{SonarCloud} Vulnerable Code Prospector for C) is an online tool 
for collecting vulnerable source code from open-source repositories linked to SonarCloud. 
The tool performs static analysis and labels the potentially vulnerable source code at the file level~\cite{raducu2020:collecting}.
The Devign~\cite{zhou2018:devign} dataset includes four real-world open-source C/C++ projects,
Linux, FFmpeg, Qemu and Wireshark, where the labeling is performed using security-related keyword filtering.
The drawback of this dataset is also in the labeling: if a commit is believed to be vulnerable,
then all functions changed by the commit are labeled as vulnerable, which is always not true.
Moreover, the dataset does not classify the types of vulnerabilities encountered.
A threat recognized by the authors of these datasets/tools
is that the labeling is based on static analysis, 
which is known for its false positives, 
that may lead to incorrect labeling.
Our approach, as well as others discussed below, 
aims to mitigate this threat by building on actual security patches.
The Code Gadget Database (CGD)~\cite{li2018:vuldeepecker} collects two types of vulnerabilities in C/C++ programs:
buffer error vulnerability (CWE-119) and resource management error vulnerability (CWE-399).
The dataset covers 61638 code gadgets including 17725 vulnerable and 43913 non-vulnerable code gadgets.
It aims to improve the labeling quality by checking if the code slice of confirmed bugs overlaps
with a bug fix before labeling code as 'buggy'.
Recently, Zheng et al. \cite{zheng2021:d2a} published \emph{D2A}, 
which uses an approach based on differential analysis 
to label issues in the source code functions or snippets reported by static analysis tools. 
The D2A approach analyzes the commit messages to identify the likeliness of vulnerability fixes
from before-commit and after-commit versions. 
The dataset considers code from six open-source C/C++ projects:
OpenSSL, FFmpeg, httpd, NGINX, libtiff, and libav.

Vieira et al. \cite{vieira2019:reports} introduced a dataset of bug-fixing activities
from 70000 bug-fixing reports from 10 years of 55 open-source projects of \emph{Apache}
mined from \emph{Jira},\footnote{~https://atlassian.com/software/jira} a popular issue tracking system that  
captures information about software development, bugs, security vulnerabilities, new functionalities, etc.
The dataset's emphasis is on process-related information regarding issue management,
such as change metrics, fix effort, status, version, and assignee, but it does not go into code-level detail. 
Alves et al.~\cite{alves2016:software} analyze the bug-tracking systems of five open-source projects
(Mozilla, Linux Kernel, Xen Hypervisor, Httpd, and Glibc), 
looking for occurrences of CVE identifiers. 
They found 2875 security patches and used the associated code to build a dataset 
that labels the code before patching as vulnerable, and after patching as not vulnerable, 
as well as adding labels for vulnerability types and severity. 
Moreover, they compute software metrics at the file, class, and function level for the code before
and after patching to enable metrics-based vulnerability prediction research.
Similarly, Gkortzis et al.~\cite{gkortzis2018:vulinoss} present a vulnerability dataset 
correlating source code and software metrics of 8694 versions of open-source projects.

Jimenez et al.~\cite{jimenez2018:enabling} present \emph{VulData7}, a framework to collect a security vulnerability dataset
covering all reported NVD vulnerabilities of four security-critical open-source systems,
\emph{Linux Kernel, WireShark, OpenSSL, and SystemD}. 
Although focusing on a limited set of systems all written in the C programming language, 
their approach shares several characteristics with ours, 
but one main practical challenge for reusing their framework is that it is based on the analysis of XML files, 
a format that is no longer provided by the NVD.
Similarly, Ponta et al.~\cite{ponta2019:manuallycurated} curated a dataset of 1282 commit fixes to vulnerabilities
from 205 open-source Java projects obtained from NVD and project-specific web resources, 
and Fan and Nguyen~\cite{fan2020:code} curated a C/C++ code vulnerability dataset named \emph{Big-Vul} that corresponds
with code changes and CVE database entries from 2002 to 2019.

Lin et al.~\cite{lin2020:software} emphasize the need for a benchmark for evaluating 
the effectiveness of ML approaches aimed at fixing security vulnerability in source code.
The authors propose a benchmark dataset that offers labels at two levels of granularity, 
i.e., the file- and method level, 
collected from nine software projects written in the C programming language~\cite{lin2020:deep}.
The challenge with using this dataset for training purposes 
is that it contains only 1471 vulnerable functions and 1320 vulnerable files,
which is rather small for training a deep learning model.

\head{Software Repositories and Repository Mining Frameworks}
Ma et al.~\cite{ma2021:world} present World of Code (WoC), 
a large and frequently updated collection of git-based version control data. 
The data is indexed three storage abstractions and arranged in four layers.
New and updated repositories are periodically analyzed, and the database is updated once a month.
The dataset stores four types of raw git objects: commits, trees, blobs and tags.
The bulk extraction of these raw git objects was done in parallel using \emph{libgit2},
a portable tool implemented in C.
Querying this database can help answer different research questions, 
such as trends in programming languages used, ecosystem comparisons, bug prediction,
developer migration studies, bot detection, understanding developer trajectories, etc.
In the same way, the dataset can be inspected for different security-related research areas,
vulnerable commits classification, developers' tendency towards vulnerability, etc.

\emph{Perceval}~\cite{duenas2018:perceval} is an automatic and incremental data gathering tool
to mine software development data from various sources such as versioning systems,
bug tracking systems and mailing lists.
The JSON output of the tool stores commit-level information of the repositories that can be used
to analyze the activities of the repositories and the developers.
However, the tool does not expose access to the finer granularity levels of open-source projects, 
such as the code-level vulnerability information that we are concerned with.

\emph{PyDriller}\footnote{~https://github.com/ishepard/pydriller}
is a Python package to mine code-level information from different version control systems~\cite{spadini2018:pydriller}.
It allows extracting commit-, file-, method levels information from git repositories, 
and it collects all relevant meta-information, including the computation of a number of code-level metrics.
Because of these features, we use \emph{PyDriller} during the construction of \CVEfixes, 
before mapping the collected data to a relational database.

\head{Differences}
The \CVEfixes dataset presented in this paper differs from the existing datasets in the following ways:
(1) In general, it covers more information about the vulnerability at different levels of granularity, 
    up to the actual source code before and after fixing.
(2) It covers a longer timeframe, being based on all CVE records available at the collection date of \jsonPullDate.
Moreover, it's not restricted to this timeframe: at any point in the future,
our publicly available tool can automatically collect and integrate the data 
up to the last published CVE at that point in time.
(3) Unlike many other datasets that focus on specific programming languages like Java or
C/C++, our dataset covers multiple programming languages and allows the language as a query attribute.
(4) Some of the existing vulnerability datasets only classify the source code as vulnerable or not.
In addition to this binary classification, 
\CVEfixes classifies the vulnerabilities into categories as defined by Common Weakness Enumeration (CWE) types, 
as well as CVSS severity scores,
enabling research into multi-class vulnerability prediction.

\section{Dataset Construction}
\label{sec:construction}

\subsection{Overall Workflow}

\noindent
The National Vulnerability Database (NVD)~\cite{nist:nvd} is a repository of
vulnerability management data maintained by the National Institute of Standards and Technology (NIST).
NIST publishes the entire NVD database for public use employing various data feeds, 
including JSON vulnerability feeds, security checklist references, 
security-related software flaws, misconfigurations, product names, and impact metrics.
The JSON vulnerability feed comprises a daily updated feed of vulnerability records (CVEs), 
organized by year of origin (updates may concern CVEs published in earlier years).

Each vulnerability in the file has its CVE-ID, publish date, description, associate reference
links, vulnerable product configuration, CWE weakness categorization~\cite{mitre:cwe}, and other metrics.
Moreover, each vulnerability's severity is ranked using
the Common Vulnerability Scoring System (CVSS)~\cite{first:common}.
The CVSS is composed of a set of metrics in three groups, i.e., base, temporal, and environmental, 
to specify the characteristics and contextual information of a vulnerability.
There are two versions of the scoring system: CVSS$_\text{v2}$
and CVSS$_\text{v3}$, that are both in active use in the NVD.\footnote{~https://nvd.nist.gov/vuln-metrics/cvss}
CVSS$_\text{v3}$ introduces a number of changes over CVSS$_\text{v2}$ to score vulnerabilities more accurately 
and provide more information to distinguish between different types of vulnerabilities.
Finally, when a vulnerability in an open-source project is fixed, 
the CVE record will be updated with one or more pointers to the relevant source code repositories,
as well as commit hashes of the fixes.
These comprehensive, scrutinized, and frequently updated JSON vulnerability feeds 
are the primary basis for generating the vulnerability data in \CVEfixes.

\begin{figure}[t]
    \vspace*{1ex}
    \centering%
    \includegraphics[width=0.85\columnwidth,trim=0 0 43 0,clip]{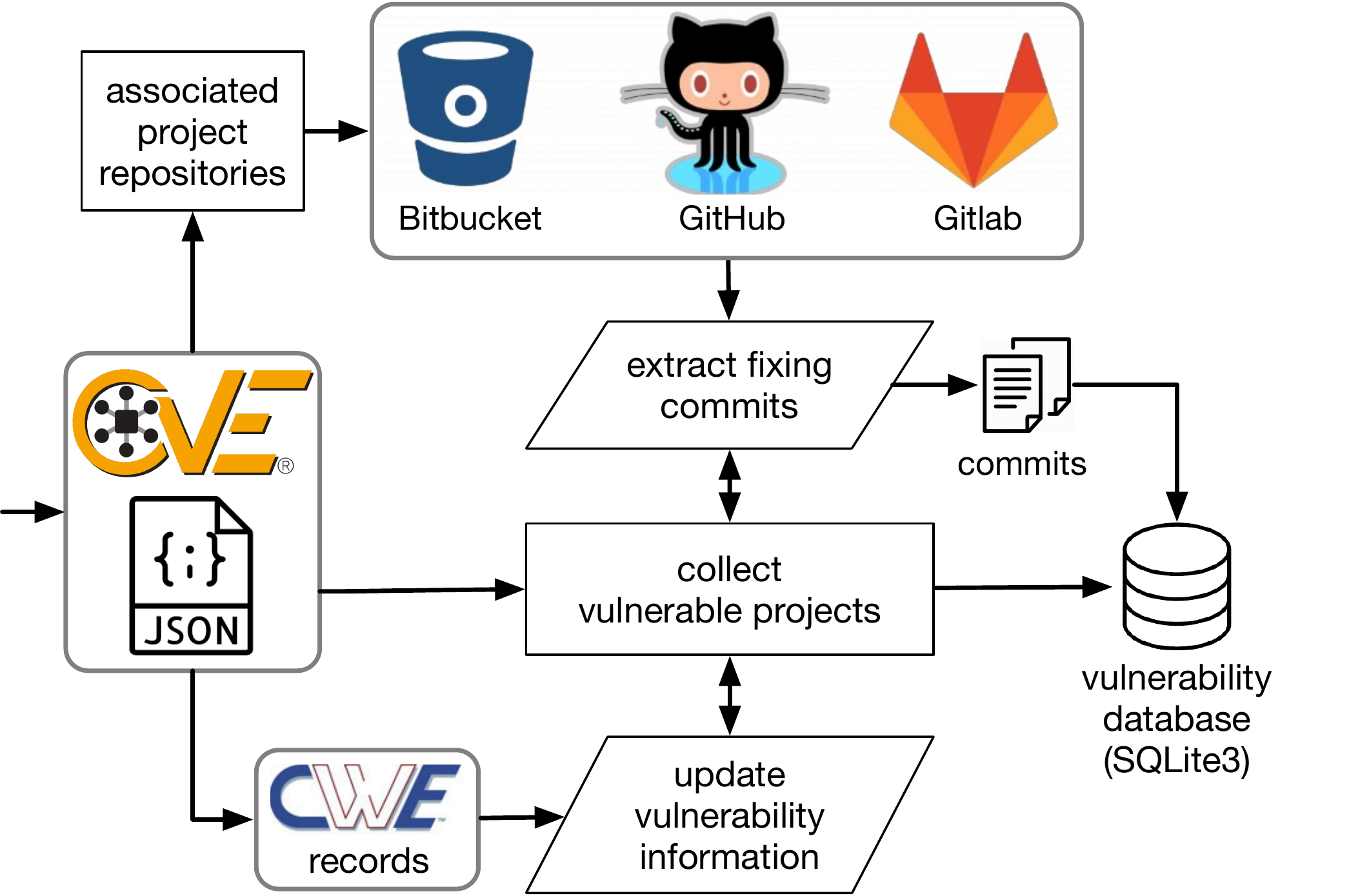}
    \vspace*{-1ex}
    \caption{Dataset construction workflow}
    \label{figure:dataset_construct}
\end{figure}

Figure~\ref{figure:dataset_construct} presents the overall workflow of the dataset collection process.
It starts by collecting CVE records via JSON vulnerability feeds.
Since these CVE vulnerabilities are categorized according to CWE weakness types~\cite{mitre:cwe}, 
we also collect the details of these CWE types and cross-reference them with the appropriate CVE records.

Moreover, for CVEs that have fixes associated with them, 
the corresponding open-source project repositories are (temporarily) locally cloned, 
and source-level information about the vulnerable code and the corresponding fixes is gathered based on the commit hashes reported in the CVE record.
After extracting the vulnerable and fixed code,
a number of code-level metrics are computed, and additional meta-data such as the programming language is derived. 
Finally, the collected information is stored in a relational database.

\begin{figure*}[t]
    \vspace*{1ex}
    \centering
    \includegraphics[width=0.8\textwidth]{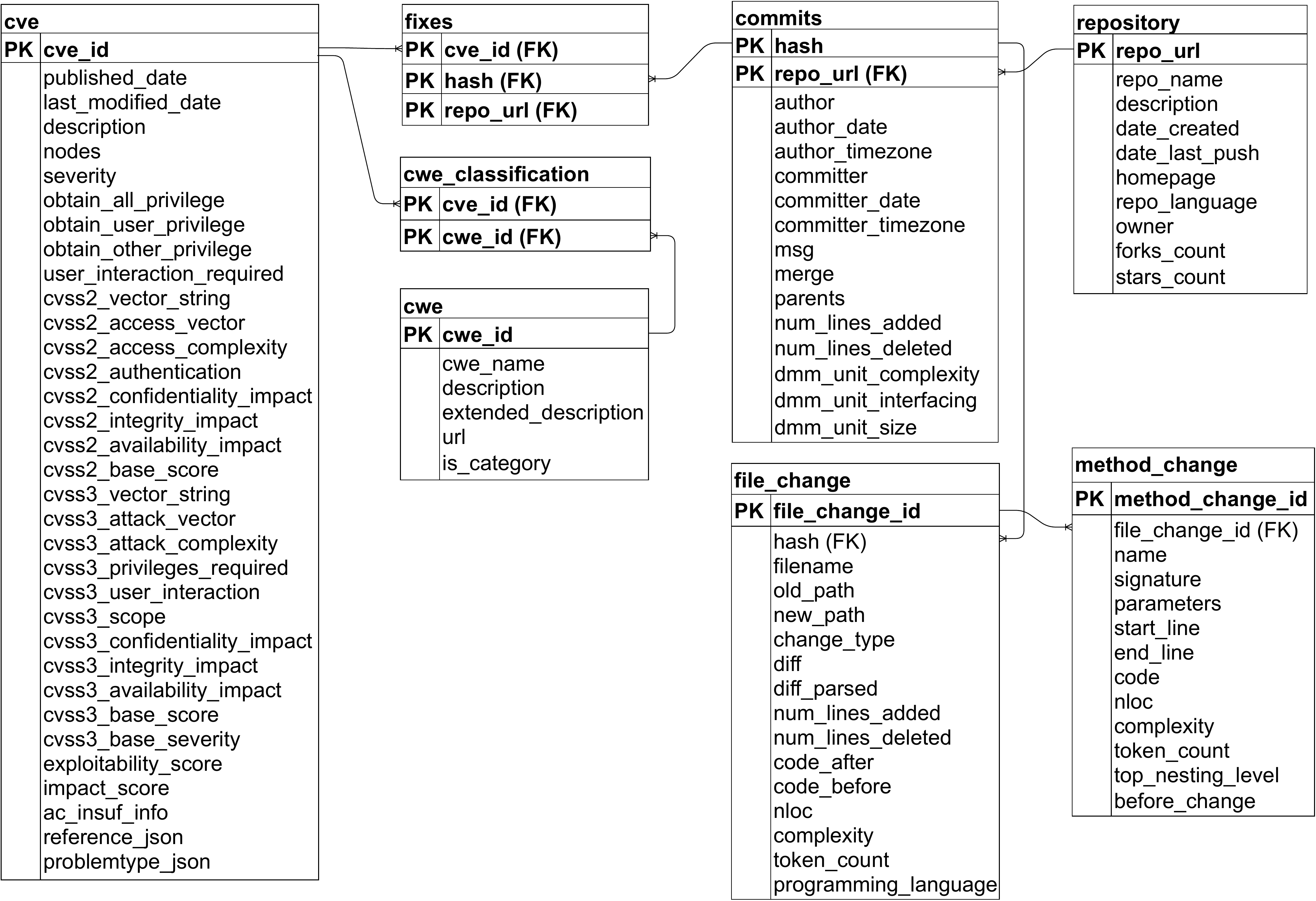}
    \vspace*{-.5ex}
    \caption{Entity-Relationship Diagram showing the various levels of abstraction in \CVEfixes and their interconnections}
    \label{Figure:ERD}
\end{figure*}

The overall structure of the database is shown in Figure~\ref{Figure:ERD}. 
The main tables capture information about CVE records, CWE records, OSS repositories, commits, changed files, and changed methods. 
Two additional tables respectively associate fixes with CVEs, and CWE classifications with CVEs.

\subsection{Details of the Automated Collection Tool}

The proposed process has been implemented in a fully automated \CVEfixes collection tool, 
and the remainder of this section discusses the major phases implemented in the tool.
The collection can easily be repeated to keep up-to-date with newly discovered or patched vulnerabilities. 
Since updates can affect any of the published CVEs,
the collection tool currently revisits all published CVEs to check if relevant fixes were added. 
One direction for future work is enabling a more incremental process.

\head{Scanning of CVE Records:}     %
By default, the collection tool retrieves all published JSON vulnerability feeds from the NVD server, 
covering the first published CVE in 2002 up to the last published one on the date of collection (the feeds are updated daily).
The JSON files are then aggregated, flattened, 
and processed to filter out redundant information from the CVE records 
(in the original feeds, some fields are repeated at deeper nesting levels).
All CVE records that do not have fixes associated with them in the \emph{reference\_json} field are ignored,
because gathering corresponding vulnerable or fixed code is not possible for these cases.
After processing and filtering the records, various details of the vulnerability such as
CVE-ID, published date, last modified date, reference data,
CVSS severity scores, vulnerability impact and scope,
exploitability score, are recorded in the \ERtable{cve} table.

\begin{sloppypar}
\head{Classification of Vulnerabilities:}   %
As discussed before, CVE records in the NVD are annotated with CWE vulnerability types 
that indicate at a more abstract level what kind of vulnerability this CVE concerns. 
The \emph{problemtype\_json} field of the NVD JSON vulnerability feeds may refer to one or more of these CWE types.
The NVD only uses a subset\footnote{~https://nvd.nist.gov/vuln/categories} of the full CWE list maintained by MITRE,\footnote{~\label{cwefn}https://cwe.mitre.org/data/index.html} 
and does not distinguish between individual CWEs and CWE categories, referring to either as ``CWE-\emph{<num>}''.
Moreover, two additional classifications, ``\emph{NVD-CWE-noinfo}'' and ``\emph{NVD-CWE-other}'',
are used to refer to cases where, respectively, there is not enough info for classification, 
or where the NVD deviates from the full CWE.
Note that we also found around 250 cases where the NVD feeds contain the CWE classification ``\emph{unknown}''.
For consistency, we map these ``\emph{unknowns}'' to ``\emph{NVD-CWE-noinfo}'' in our dataset.
\end{sloppypar}

\head{Meta-Information for Repositories:}   %
At the time of writing, the majority (>80\%) of fixes in the NVD refer to code on one of the three major OSS forges, GitHub, GitLab, and Bitbucket, 
with 98\% of those being on GitHub. 
The remaining 20\% of fixes point to other forges (e.g., sourceforge or the defunct gitorious), 
or to project-specific servers hosting other versioning systems (e.g., mercurial, subversion, or CVS).
As a result, we focus on gathering code from those three OSS forges,
and we have modeled the \ERtable{repository} table to maximally cover GitHub's repository meta-information, 
such as repository name, description, date of creation, last date of push, homepage, programming language,
number of forks and number of stars.
This information can be used to, for example, focus on certain programming languages 
or filter for popular repositories by setting a minimum threshold on the number of stars.
The latter is also a common and effective approach to focus on information related to \emph{mature} projects.
Note that GitLab and Bitbucket do not offer all of these attributes, 
so clearly distinguishable values are used to represent missing information,
such as the value ``-1'' for the \emph{stars\_count} of Bitbucket repositories (which does not offer stars).

\head{Extraction of Commits:} %
Whenever an OSS vulnerability is fixed, 
the \emph{reference\_json} in the CVE record contains the repository URL,
as well as a pointer to the exact commit that introduced the fix, by means of a git hash.
We use this information to locally (and temporarily) clone the repository, 
and use it to extract versions of the code before and after submitting the fix.
We extract this information at the commit-, file- and method level, 
where each entry in the \ERtable{commits} table is associated to one or more \ERtable{cve}s via the \ERtable{fixes} table.
In addition, the \ERtable{commits} table stores meta-information about the commit, 
such as the author, time and date, commit message, if it is a merge commit 
(often to include pull requests in projects using GitHub Flow),
the number of lines added or deleted, 
and Delta Maintainability Model metrics related to the change~\cite{biase2019:delta}.

\head{Extracting the Modified Files:}   %
Every entry in the \ERtable{commits} table is linked via the commit hash to one or more \ERtable{file\_change}s 
that were included to fix the vulnerability (or vulnerabilities) associated with the commit.
For each file change, the dataset contains the contents of the file before and after making the change, 
in \emph{code\_before} and \emph{code\_after} respectively,
as well as the \emph{diff} of the change in the format delivered by Git, 
and a parsed version of this information in \emph{diff\_parsed}, 
containing a dictionary of added and deleted lines. 
In addition, several meta-data are collected, 
such as filename, old and new path, type of the modification (i.e., added, deleted, modified, or renamed), 
number of lines added or removed in that file, lines of code (\emph{nloc}) after the change, 
and cyclomatic complexity after the change.
Finally, the \emph{Guesslang} tool\footnote{~https://guesslang.readthedocs.io/en/latest/}
is used to detect the actual \emph{programming\_language} that is used in a given file.
\emph{Guesslang} is an open-source deep-learning based classifier that was trained with over a million source code files,
and recognizes over 30 programming languages. 
Note that these detected languages may differ from, 
and are more precise than the repository language that is, for example, reported by GitHub.

\head{Extracting the Modified Methods:}     %
Similar to the changes at the file-level, 
we keep track of changes at the method-level in the \ERtable{method\_change} table.
In addition to the actual code, 
each method change stores meta-information such as the method name, 
its signature, parameters, start and end line of the method,
lines of code, cyclomatic complexity, 
and a boolean indicating whether this concerns the code from before committing the fix or not.

\subsection{Practical Guidance}

\noindent
The automated \CVEfixes dataset collection tool is distributed via
GitHub.\footnote%
{~\url{https://github.com/secureIT-project/CVEfixes}, \textsc{doi:}\href{https://doi.org/10.5281/zenodo.5111494}{10.5281/zenodo.5111494}.}
The distribution comes with detailed requirements and installation instructions. 
It requires Python3.8 or later to run, and SQLite3 to construct and store the data into a relational database.
The main Python packages that it depends on are pandas, numpy, requests, 
PyDriller, PyGithub and guesslang.
We provide \emph{requirements.txt} and \emph{environment.yml} files for dependency management using pip/venv or anaconda/miniconda.

The tool is configured by means of a number of settings in a \emph{.CVEfixes.ini} file (an example is provided in the distribution): 
\begin{itemize}
    \item \emph{database\_path}: location of the \CVEfixes database file (also used to hold some temporary files during extraction).
    \item \emph{sample\_limit}: an optional limit on the number of commits to be gathered (mainly for testing and demonstration purposes, sample\_limit = 0 means unlimited collection).
    \end{itemize}
Note that the GitHub API is severely rate-limited when unauthenticated access is used.
These limits can be raised significantly to up to 5000 requests per hour 
by authenticatinsg with a username and personal access token.\footnote{~https://docs.github.com/en/github/authenticating-to-github/creating-a-personal-access-token}
A \emph{github\_username} and \emph{github\_token} can be configured in \emph{.CVEfixes.ini} for this purpose. 
With a sample limit of 25, no token is needed and rate-limiting will not be triggered.

We provide a compressed SQL dump for the initial release of the \CVEfixes vulnerability dataset that covers
all published CVEs in the NVD up to \jsonPullDate at Zenodo.%
\footnote{~\url{https://zenodo.org/record/4476563}, \textsc{doi:}\href{https://doi.org/10.5281/zenodo.4476563}{10.5281/zenodo.4476563}.}
The distribution contains a simple shell script to convert the compressed SQL dump into an SQLite3 database:
\begin{lstlisting}
    $ sh Code/create_CVEfixes_from_dump.sh
\end{lstlisting}
Alternatively, the \CVEfixes dataset can be gathered from scratch using the following shell script: 
\begin{lstlisting}
    sh Code/create_CVEfixes_from_scratch.sh
\end{lstlisting}
Note that this process will overwrite an existing database. 
Moreover, at the time of writing, 
the full collection process can take up to 15 hours, 
depending on the available internet connection.
The advantage of taking this route is that the database will contain all CVEs published up to the date of initiating the collection process.

The distribution contains an example Jupyter Notebook \emph{Examples/query\_CVEfixes.ipynb} 
that shows how to explore the \CVEfixes dataset 
and contains the code to generate the various tables and graphs in this paper.

Recall the overall structure of the database shown in Figure~\ref{Figure:ERD}.
Then the following query extracts all methods involved in fixes related to C programs before the changes were made:
\begin{lstlisting}
    SELECT m.name, m.signature, m.nloc, m.parameters, m.token_count, m.code
    FROM method_change m, file_change f
    WHERE f.file_change_id = m.file_change_id
    AND f.programming_language = 'C'
    AND m.before_change = True
\end{lstlisting}
Another example, inspired by the ManySStuBs4J dataset~\cite{karampatsis2020:how}, to find all files that only add or remove a single line as 
fix:
\begin{lstlisting}
    SELECT cv.cve_id, f.filename, f.num_lines_added, f.num_lines_deleted, f.code_before, f.code_after, cc.cwe_id
    FROM file_change f, commits c, fixes fx, cve cv, cwe_classification cc
    WHERE f.hash = c.hash AND c.hash = fx.hash
    AND fx.cve_id=cv.cve_id 
    AND cv.cve_id=cc.cve_id
    AND f.num_lines_added<=1
    AND f.num_lines_deleted<=1;
\end{lstlisting}

\section{Dataset Exploration}
\label{sec:statistics}

\noindent
Table \ref{Table:dataset_summary} presents the summary statistics of the initial release of the \CVEfixes dataset.
This initial release covers 
\cveCount unique CVEs in \projectCount OSS projects, %
with \commitCount unique vulnerability fixing commits. %
The CVEs are classified into \cweCount different CWE vulnerability types.
Note that some of the vulnerability fixing commits cover multiple CVEs,
as is indicated by the number of commits being lower than the number of CVEs.

\begin{table}[b]
  \vspace*{-1ex}%
  \centering%
  \caption{Summary statistics of the \CVEfixes dataset}
  \label{Table:dataset_summary}
  \vspace*{-1ex}%
  \tablefont%

\pgfplotstabletypeset[
  columns/methods/.style={fixed},
  column type=,
  begin table={\begin{tabular}{cccccc}},
  end table={\end{tabular}},
  outfile={results/dataset_summary.pgfplotstable},
  ]{results/dataset_summary.csv}
\end{table}

\begin{table}[t]
  \vspace*{1ex}
  \centering%
  \caption{Top 10 projects in \CVEfixes with respect to
    (a) the number of fixed CVEs, shown with the project's security metrics,
    (b) number of vulnerability fixing commits,
    and the number of (c) files and (d) methods involved in these fixes.}
  \label{Table:projects_list}
  \vspace*{-1ex}%
  \tablefont%
  \hspace*{2pt}%
\pgfplotstabletypeset[
    every head row/.style={
	    before row={\toprule & & \textbf{average} & \textbf{average} & \textbf{average} & \textbf{average} \\},
		  after row={\midrule},
    },
  columns={projects,cve_count,avg_cvss_v2_score,avg_cvss_v3_score,avg_exploitability_score,avg_impact_score},
  sort,sort key=cve_count, sort cmp=float >,
  columns/projects/.style={string type, column name=\textbf{project}},
  columns/cve_count/.style={column name=\textbf{\#\,CVEs}},
  columns/avg_cvss_v2_score/.style={column name=\textbf{CVSS$_\text{v2}$},fixed zerofill,precision=2},
  columns/avg_cvss_v3_score/.style={column name=\textbf{CVSS$_\text{v3}$},fixed zerofill,precision=2},
  columns/avg_exploitability_score/.style={column name=\textbf{exploitability},fixed zerofill,precision=2},
  columns/avg_impact_score/.style={column name=\textbf{impact},fixed zerofill,precision=2},
  column type=,
  skip rows between index={10}{9999},
  begin table={\begin{tabularx}{0.975\columnwidth}{@{}Xrrrrr@{}}},
  end table={\end{tabularx}},
  outfile={results/projects_list1.pgfplotstable},
]{results/severity_summary.csv}\newline%

(a)\vspace*{2ex}

\begin{filecontents*}{results/vul_projects_10.csv}
commit_project,commit_count,file_project,file_count,method_project,method_count
linux,1029,linux,2007,GeniXCMS,14235
ImageMagick,171,GeniXCMS,1930,linux,4303
tensorflow,156,kanboard,698,exponent-cms,3048
phpmyadmin,126,CycloneTCP,611,Ilch-2.0,1427
tcpdump,101,X2CRM,594,arangodb,1224
FFmpeg,88,exponent-cms,445,kanboard,1169
php-src,61,Ushahidi-Web,402,tensorflow,1045
MISP,54,wityCMS,372,moped,950
WordPress,47,tcpdump,344,WordPress,915
radare2,43,tensorflow,340,X2CRM,790
\end{filecontents*}
\pgfplotstabletypeset[
columns={commit_project,commit_count},
columns/commit_project/.style={string type, column name=\textbf{project}},
columns/commit_count/.style={column name=\textbf{\#\,commits}},
column type=,
begin table={\begin{tabularx}{0.30\columnwidth}{@{}X@{}r@{}}},
end table={\end{tabularx}},
outfile={results/projects_list2.pgfplotstable},
]{results/vul_projects_10.csv}
\mbox{~~}
\pgfplotstabletypeset[
columns={file_project,file_count},
columns/file_project/.style={string type, column name=\textbf{project},
  string replace={Ecommerce-CodeIgniter-Bootstrap}{Ecommerce*} %
  },
columns/file_count/.style={column name=\textbf{\#\,files}},
column type=,
begin table={\begin{tabularx}{0.28\columnwidth}{@{}X@{}r@{}}},
end table={\end{tabularx}},
outfile={results/projects_list3.pgfplotstable},
]{results/vul_projects_10.csv}
\mbox{~~}
\pgfplotstabletypeset[
columns={method_project,method_count},
columns/method_project/.style={string type, column name=\textbf{project},
  string replace={angular-expressions}{angular-expr} %
  },
columns/method_count/.style={column name=\textbf{\#\,methods}},
column type=,
begin table={\begin{tabularx}{0.35\columnwidth}{@{}X@{}r@{}}},
end table={\end{tabularx}},
outfile={results/projects_list4.pgfplotstable},
]{results/vul_projects_10.csv}\newline%

(b) \hspace*{.26\columnwidth} (c) \hspace*{.3\columnwidth} (d) \mbox{~~}

\vspace*{-3ex}
\end{table}

To further characterize the data, Table \ref{Table:projects_list} presents the top ten projects in \CVEfixes with respect to the number of CVEs,
number of vulnerability fixing commits, and number of files and methods involved in fixes.
In particular, Table~\ref{Table:projects_list}a corresponds with the project-wise number of CVEs,
the average CVSS$_\text{v2}$ and CVSS$_\text{v3}$ base scores of these vulnerabilities,
as well as their average exploitability, and impact scores.
Not surprisingly, the long-running and well-scrutinized \cveFirst project has the most vulnerabilities and vulnerability fixing commits
(shown in resp. Table~\ref{Table:projects_list}a and \ref{Table:projects_list}b),
i.e., \commitFirstCount commits for fixing \cveFirstCount vulnerabilities.
Note that \cveFirst has roughly six times as much vulnerabilities as \cveSecond,
the project with second most vulnerabilities (\cveSecondCount),
yet the security impact of those discovered in \cveSecond is considerably larger,
as shown by the various security metrics.

Looking at the projects in the four top ten lists of Table~\ref{Table:projects_list}a-\ref{Table:projects_list}d,
it is not surprising to see some overlap between the lists.
On the other hand, it \emph{is} surprising to see that some projects,
such as \emph{GeniXCMS}, \emph{kanboard}, and \emph{exponent-cms},
which do not occur at all in the top ten CVEs and \#commits,
come in very highly ranked on the amounts of files and methods that need to be changed to fix vulnerabilities in these projects.
This sudden rise in the ranks suggests it can be of interest to investigate
how the modularization decisions in these projects impacted the security fixes that were made,
and may indicate an application-level code smell such as shotgun surgery.

This finding also shows that there is \emph{no} direct correlation between the number of files or methods and the number of CVEs (or commits to fix them),
because that would mean the respective top tens would have been similar.
We \emph{do} see the expected relation between CVEs and vulnerability fixing commits
(with minor changes due to some commits covering multiple CVEs, as noted earlier).

The 10 projects in Table~\ref{Table:projects_list}a cover \percentCveTen of the total number of CVEs,
those in Table~\ref{Table:projects_list}b cover \percentCommitTen of the total number of vulnerability fixing commits,
the ones in Table~\ref{Table:projects_list}c account for \percentFileTen of the total number of files involved in fixes,
and finally, the ten projects in Table~\ref{Table:projects_list}d make up \percentMethodTen of the total number of methods involved in fixes.

Table~\ref{Table:cwe_summary} presents the ten most occurring CWE types based on CVE count.
We see that CWE-79 (Improper Neutralization of Input During Web Page Generation,
is the most commonly identified vulnerability type.
This type is commonly known as Cross-site Scripting or XSS.
It has been assigned to 635 CVEs and was fixed in 670 commits that changed 3226 files.
Observe that the number of vulnerability fixing commits is larger than the number of CVEs, 
which is caused by the fact that some of the CVEs are associated with multiple fixes. 
This occurs, for example, when a CVE is fixed in several related projects, 
or when the fix is spread over multiple commits.
On second place comes 
CWE-119 (Improper Restriction of Operations within the Bounds of a Memory Buffer),
reported in 408 CVEs and fixed in 403 commits that changed 716 files.

Next, we find CWE-20 (Improper Input Validation) assigned to 382 CVEs.
which is closely followed by CWE-125 (Out-of-bounds Read),
which is actually a ``child'' of the more general CWE-119 category,
and indicates that an index pointer is used beyond the bounds of the memory buffer.
It is reported in 380 CVEs and fixed in 404 commits that modified 1159 files.
After this top four, we see a considerable drop in the number of CVEs assigned to lower-ranked CWEs.
The top ten CWEs cover approximately
\percentCveCweTen of the total number of CVEs,
\percentCommitCweTen of the total number of commits,
and \percentFileCweTen of the total number of files.

\begin{table}[t]
  \vspace*{1ex}
\centering
\caption{Distribution of vulnerabilities over CWE types}
\label{Table:cwe_summary}
\vspace*{-1ex}%
\tablefont%
\begin{filecontents*}{results/cwe_summary.csv}
CWE,description,cve_count,commit_count,file_count
CWE-79,Improper Neutralization of Input During Web Page Generation (Cross-site Scripting),635,670,3226
CWE-119,Improper Restriction of Operations within the Bounds of a Memory Buffer,408,403,716
CWE-20,Improper Input Validation,382,397,1965
CWE-125,Out-of-bounds Read,380,404,1061
CWE-200,Exposure of Sensitive Information to an Unauthorized Actor,276,314,818
CWE-787,Out-of-bounds Write,205,211,537
CWE-476,NULL Pointer Dereference,195,198,334
NVD-CWE-noinfo,Insufficient Information,193,219,664
NVD-CWE-Other,Other,165,170,395
CWE-264,Permissions Privileges and Access Controls,143,148,457
\end{filecontents*}
\pgfplotstabletypeset[
columns={CWE,description,cve_count,commit_count,file_count},
columns/CWE/.style={string type, column name={CWE},
  string replace={NVD-CWE-noinfo}{noinfo*},
  string replace={NVD-CWE-Other}{other*}
},
columns/description/.style={string type},
columns/cve_count/.style={string type, column name={CVEs}},
columns/commit_count/.style={string type, column name={cmts}},
columns/file_count/.style={string type, column name={files}},
column type=,
begin table={\begingroup\setlength{\tabcolsep}{5pt}\begin{tabularx}{0.99\columnwidth}{@{} m{1.35cm} @{} X @{} r @{\hspace*{3pt}} r @{\hspace*{3pt}} r @{}}},
end table={\end{tabularx}\endgroup},
outfile={results/cwe_summary.pgfplotstable},
]{results/cwe_summary.csv}
* abbreviated from NVD-CVE-noinfo and NVD-CWE-other for space.
\end{table}

\begin{table}[b]
  \centering%
  \caption{Average number of days between CVE publication and \mbox{vulnerability} fix for all CVEs in \CVEfixes.}
  \label{Table:daysFixSummary}
  \tablefont%
  \begin{filecontents*}{results/days_fix_summary.csv}
mean,std,min,25p,50p,75p,max
171.55,414.93,0.0,7.0,28.0,123.25,7636.0
\end{filecontents*}
\pgfplotstabletypeset[
  columns/methods/.style={fixed},
  columns/{25p}/.style={string type, column name={25\%}},
  columns/{50p}/.style={string type, column name={50\%}},
  columns/{75p}/.style={string type, column name={75\%}},
  column type=,
  begin table={\begin{tabular}{cccccccc}},
  end table={\end{tabular}},
  outfile={results/daysFixSummary.pgfplotstable},
  ]{results/days_fix_summary.csv}
\end{table}

\begin{table}[t]
  \vspace*{1ex}
  \centering
  \caption{Average number of days between CVE publication and vulnerability fix for the top ten projects with most CVEs (cf.\ Table~\ref{Table:projects_list}a).}
  \label{Table:daysToFix}
  \tablefont%
  \hspace*{2pt}\pgfplotstabletypeset[
  every head row/.style={
      before row={\toprule},
      after row={\midrule},
  },
  columns={projects, cve_count, cwe_count, min_days_to_fix, max_days_to_fix, median_days_to_fix, mean_days_to_fix},
  columns/projects/.style={string type, column name=\textbf{project}},
  columns/cve_count/.style={column name=\textbf{\#\,CVEs}},
  columns/cwe_count/.style={column name=\textbf{\#\,CWEs}},
  columns/mean_days_to_fix/.style={column name=\textbf{mean},fixed zerofill,precision=2},
  columns/median_days_to_fix/.style={column name=\textbf{median},fixed zerofill,precision=0},
  columns/max_days_to_fix/.style={column name=\textbf{max},fixed zerofill,precision=0},
  columns/min_days_to_fix/.style={column name=\textbf{min},fixed zerofill,precision=0},
  column type=,
  skip rows between index={10}{9999},
  begin table={\begin{tabularx}{0.95\columnwidth}{@{}Xrrrrrr@{}}},
  end table={\end{tabularx}},
  outfile={results/datsToFix.pgfplotstable},
  ]{results/severity_summary.csv}
\end{table}

We also investigated if it was possible to compute how long it took to fix the vulnerabilities, 
from the day that the CVE was disclosed to the project to the day that the fix was committed. 
Unfortunately, the NVD only includes the CVE \emph{publication date}, 
which generally does not indicate when the vulnerability was discovered 
or when it was shared with the project. 
The actual disclosure date can be several months earlier if a responsible disclosure process is used.
For example, Google's Project Zero delays publication for 90 days after disclosing vulnerabilities that are not actively being exploited,
giving the option for earlier publication with mutual agreement.
Other security researchers may use other time frames or agree on them on a case-by-case basis with the projects.
As a result of the variability this introduces in the time between disclosure and publication time, 
it is impossible to make strong statements on how long the projects really used to address the vulnerability.

With the above caveat in mind, Table~\ref{Table:daysFixSummary} shows some statistics for the complete \CVEfixes dataset 
on how many days it took from the CVE publication to committing a fix for the vulnerability.
Moreover, Table~\ref{Table:daysToFix} presents such statistics for the top ten projects with most CVEs, 
as earlier presented in Table~\ref{Table:projects_list}a.

The last two characteristics that we will investigate are the vulnerability severity scores of CVEs in \CVEfixes,  
as well as the impact that their fixes had on the maintainability of the project by means of the delta maintainability model (DMM).
The delta maintainability model (DMM)~\cite{biase2019:delta} 
is a measure to compare and rank fine-grained code changes
into low and high-risk changes.
The overall DMM score refers to the proportion of low-risk changes in a commit, 
and is computed as the mean of three individual metrics that respectively measure 
the risk associated with code size, (cyclomatic) complexity, and size of the interface.
The values of each of the DMM metrics range from 0 to 1,
with higher values indicating better maintainability, i.e., lower risk changes, 
and low values indicating poor maintainability, i.e., high risk changes.

\begin{figure}[b]
  \centering
  \includegraphics[width=0.9\columnwidth]{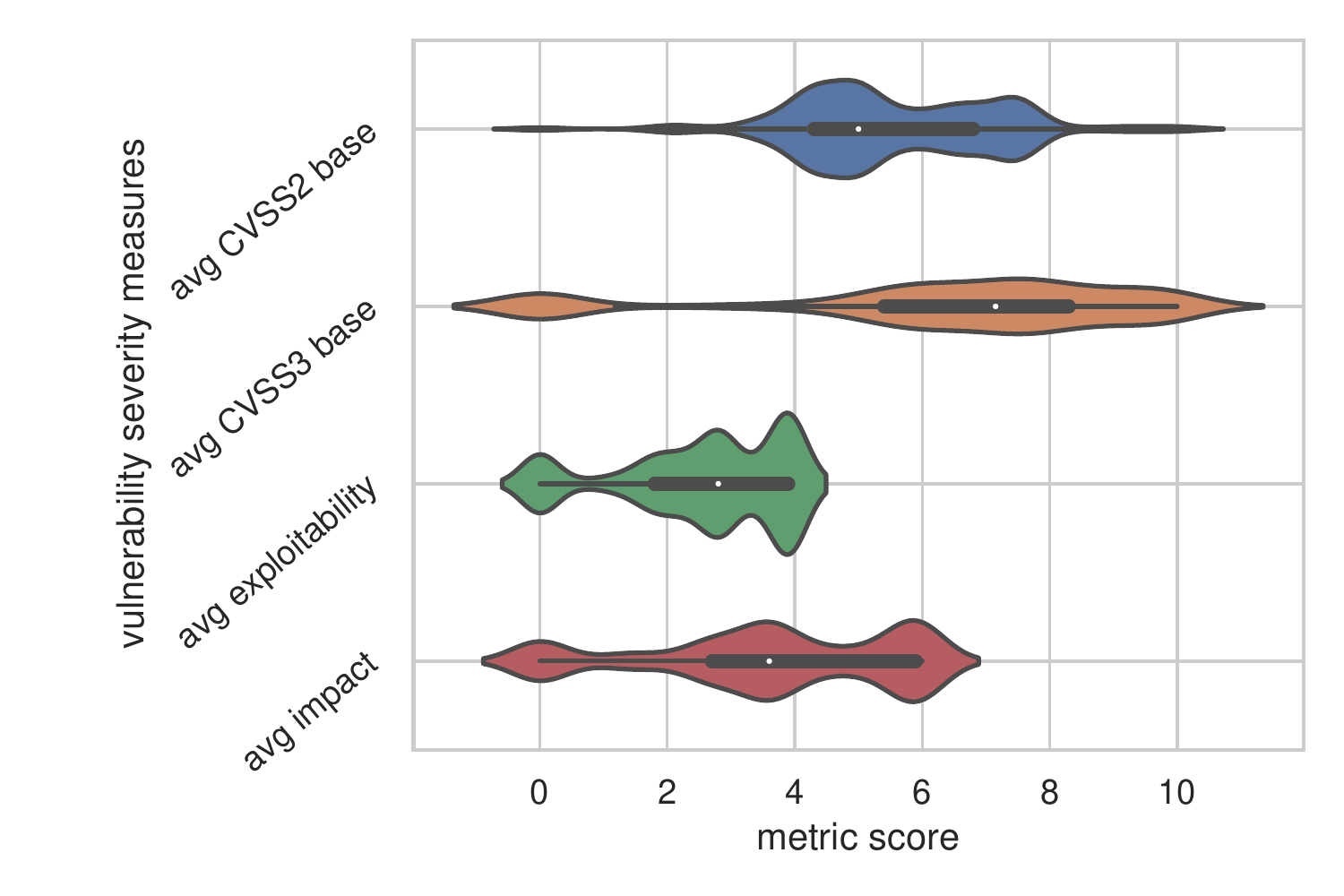}
  \vspace*{-1.5ex}
  \caption{Violin plot showing the distribution of average vulnerability severity scores for projects included in \CVEfixes}
  \label{figure:severityViolin}
\end{figure}

Figure~\ref{figure:severityViolin} presents violin plots that show the distribution of 
CVSS$_\text{v2}$ and CVSS$_\text{v3}$ base scores,
as well as exploitability and impact, 
for all vulnerabilities in \CVEfixes, aggregated into averages per project.
The width of each violin corresponds to the frequency of data points with that average. 
Inside the violin there is small box plot showing the ends of the first and third quartiles, 
and the median is indicated by the white dot.
The values of these vulnerability severity metrics can vary between 0 and 10, with lower being better, and higher being worse.
The figure suggests that the majority of the vulnerabilities have severity scores that are on the high side of the severity range, 
although their exploitability and impact tends to be on the lower sides of their respective ranges.

\begin{figure}[t]
  \vspace*{1ex}
  \centering
  \includegraphics[width=0.9\columnwidth]{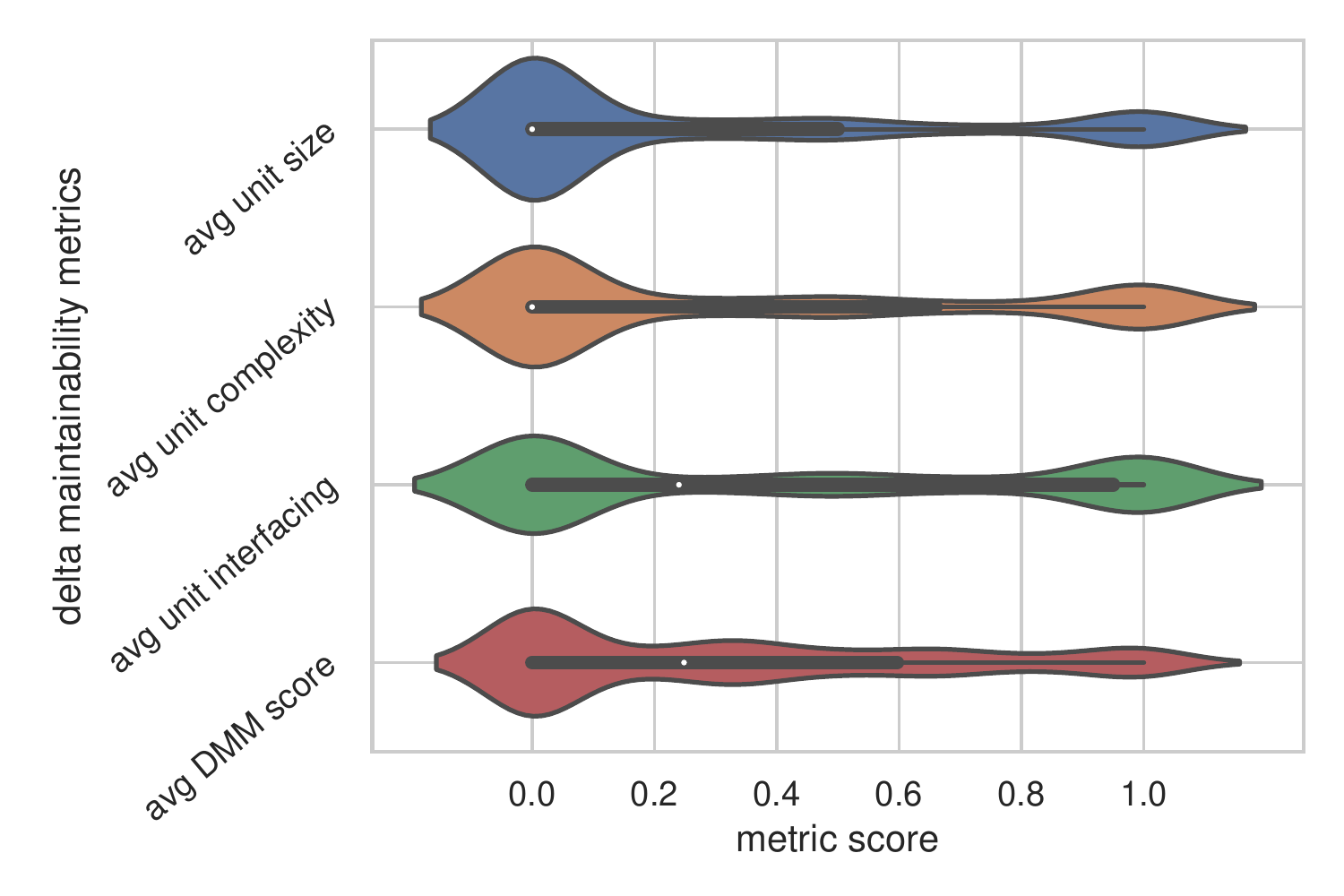}
  \vspace*{-1.5ex}
  \caption{Violin plot showing the distribution of average DMM scores for fixes to the projects included in \CVEfixes}
  \label{figure:dmm_plot}
\end{figure}

Figure~\ref{figure:dmm_plot} presents violin plots that show the distribution of 
DMM metrics of all vulnerability fixing commits in \CVEfixes, 
again aggregated into averages per project.
These metrics show how the maintainability of the projects was affected by the security fixes.
Observe that the individual DMM metrics \emph{unit size}, \emph{unit complexity}, and \emph{unit interfacing} mostly
have a high density around 0 (indicating high risk changes with a detrimental effect on maintainability), 
and a smaller number of changes are considered low risk (with scores closer to 1).
The overall \emph{DMM score} is an aggregate computed as the mean of the individual metrics~\cite{biase2019:delta}.
As such, it is more universally spread across different risk values, 
but still with a considerable number of commits requiring high-risk code change to fix the vulnerabilities, 
with the median around 0.25 and the third quartile around 0.6.
An interesting question for follow-up research is investigating if vulnerability fixing commits that lower maintainability 
are followed by refactoring commits that restore maintainability.

\section{Discussion}
\label{sec:using}

\subsection{Applications of \CVEfixes}

\noindent
The \CVEfixes dataset can be used in several ways to support data-driven software security research, 
for example, for automated vulnerability prediction, 
automated classification of identified vulnerabilities in Common Weakness Enumeration (CWE) types,
vulnerability severity prediction, 
and automated repair of security vulnerabilities.
The remainder of this section discusses several of these applications in more detail.

\head{Automated Vulnerability Prediction/Identification}
The \CVEfixes dataset contains different levels of vulnerability data, 
such as CVE-ID, version, description, type, publication date, current status, 
all interlinked up to the actual code changes that were introduced to fix the vulnerability.
This data can be used to extract code features and metrics that help 
to better understand how security vulnerabilities are introduced in code.
Moreover, the features and metrics can be used to model, train, and test vulnerability predictors 
based on classical machine learning approaches, 
and the textual descriptions, ranging from the CVE vulnerability level to the code level, 
can be used to train deep learning-based models for vulnerability prediction.
Many studies have already targeted automated vulnerability identification using various machine learning models~\cite{li2020:automated, russell2018:automated, saccente2019:project, dam2018:automatic}, 
and Section~\ref{sec:relwork} surveyed some of the challenges with the datasets used in these studies, 
such as limited size, lack of representativeness, and dataset imbalance. 
A comprehensive dataset like \CVEfixes helps to overcome these challenges and enables a more thorough evaluation of the approaches.

\head{Automated Vulnerability Classification}
The inclusion of vulnerability classifications in \CVEfixes allows us to go one step further than automated vulnerability prediction, 
it also enables research on automated vulnerability classification, 
i.e., not just predicting the presence of a vulnerability but characterizing the type of the vulnerability.
Such a classification is of interest since (automated) program repair approaches may have different efficacy and efficiency for different vulnerability types, 
so knowing the vulnerability type helps to inform which repair strategy to take.
To address this challenge, $\mu$VulDeePeaker~\cite{zou2019:mvuldeepecker} and
ManySStuBs4J~\cite{karampatsis2020:how} have constructed the multi-class vulnerability dataset.
However, those datasets are specific to a given programming language and certain vulnerability/bug types.
\CVEfixes, on the other hand, uses the well-known CWE taxonomy to classify vulnerabilities in a hierarchy of vulnerability types and categories,
and covers 27 programming languages (though 9 of these have fewer than 100 files changed in the commits covered by the dataset).

\head{Analysis of vulnerability fixing patches}
Similar to how the vulnerable code in \CVEfixes can be used to better understand how security vulnerabilities 
are introduced in code and how these can be automatically predicted, 
the fixes offered by \CVEfixes can be used to analyze and build on vulnerability fixing patches.
Several studies have already initiated research analyzing such patches, 
such as the detection of patterns that can be used in automated program repair~\cite{hegedus2020:inspecting},
and the identification of security-relevant commits, also known as \emph{pre-patches}, 
as these may inadvertently leak information about security vulnerabilities before the CVE is published~\cite{yang2018:prepatch, sabetta2018:practical}.
Other research has analyzed vulnerability fixing patches to facilitate the automated transformation of patches into ``\emph{honeypots}''
that help trap malicious actors and detect if the corresponding vulnerabilities are exploited in the wild~\cite{larmuseau2018:patchsweetner}.

\head{Automated program repair}
The pairs of vulnerable and fixed code provided by \CVEfixes can be used to train machine learning-based 
automated program repair along the lines of SequenceR~\cite{chen2019:sequencer} 
and related work surveyed in Section~\ref{sec:relwork}, 
though we need to warn for a caveat w.r.t. the completeness of the patches, as discussed in more detail in the next section.
Moreover, the current state of the art in this area is constrained with respect to the length of repairs that can be learned/synthesized, 
with 50 tokens being mentioned as an upper limit for acceptable performance~\cite{chen2019:using}.
By querying into \CVEfixes data as shown in the previous subsection
(and the example Jupyter Notebook in the distribution),
\CVEfixes facilitates the extraction of specific fixes that used only single (or a few) modified lines for fixing a vulnerability.
This can be used to focus attention of the training process on cases with a lower number of modified tokens 
which are within reach of the technology, in order to help improve patch quality.
Several studies~\cite{mashhadi2021:applying,mosolygo2021:rise,madeiral2021:largescale,hua2021:effectiveness} 
have already initiated work on the prediction of bugs/errors taking single line modified statements
into account using the ManySStuBs4J dataset~\cite{karampatsis2020:how}.
Querying \CVEfixes in the way we presented earlier enables the extension of this work towards different programming languages,
and vulnerability types.

\subsection{Limitations and Future Extensions}

\noindent
One of the limitations of the current \CVEfixes collection tool is that it collects all available fixes from scratch.
We opted for this approach since any of the CVEs in the NVD may receive updates, 
so all should be inspected anyway. 
Nevertheless, it may be interesting to explore a more incremental update scheme that 
only clones a repository if not all of its fixes are already captured in the database, 
and augments the existing fixes in the other case. 
Such an approach would help to speed up the process of updating the database. 
One caveat is that we may also need to remove fixes from the database if they are no longer referenced by a CVE, 
but as we have seen, those fixes may be referred to by other CVEs as well, 
so some additional checks would be needed.

Another challenge/limitation that we have noticed
is that some of the project repositories referenced in the NVD are no longer available.
There can be multiple reasons; for example, 
the owners might have removed/changed/renamed their repositories, 
or they may have moved the repository between forges.
This makes it impossible to fetch the fixes for these repositories.
The current implementation tries to gather code from as many repositories as possible, 
and removes references to those that are unavailable to present an internally consistent view. 
However, this choice also means that we only present a subset of the information that is available in the NVD.

A last limitation is that a commit that is referenced as a vulnerability fixing commit in the NVD
can still leave (part of) the vulnerability in the source code, 
and there may be later commits that complete the fix of the vulnerability, 
or the commit may contain changes unrelated to the fix~\cite{herzig2016:impact}.
To further improve the quality of the data, 
it would be of interest for future work to analyze consecutive patches,
and select/untangle the code that together addresses the vulnerability.

Another direction for future work is upgrading our extraction tool to support mining the repositories
from other issue-tracking and version control systems, i.e., Bugzilla, Mercurial, Subversion, etc.
This will make it possible to gather vulnerability data from an even larger collection of projects.

Finally, considering the current controversy around ML models that are trained on (A)GPL licensed code 
and may possibly regurgitate some of that code as part of their operation, 
we plan to extend our datamodel with license information for the fixes included, 
so that users of \CVEfixes can make an informed choice about including or excluding certain fixes from their training data.

\section{Concluding Remarks}
\label{sec:concl}

\noindent
In this study, we propose the \CVEfixes dataset and a fully automated collection tool that fetches this vulnerability dataset
from publicly available information in the NVD and different version control systems.
The dataset consists of real-world samples of vulnerable and corresponding patched code,
their security metrics at five levels of abstraction, all linked to the CVE records.
The collected information is enriched with various security and code related metrics 
and organized into a relational database to support easy querying.

\begin{sloppypar}
The initial release of the dataset covers \commitCount vulnerability fixing commits
from \projectCount open-source projects.
This multi-level dataset establishes both qualitative and quantitative opportunities for
vulnerability-related investigations.
Researchers can study the relation from published CVEs all the way down to the corresponding code level vulnerabilities and their proposed fixes.
The \CVEfixes dataset enables research on vulnerability detection and classification,
vulnerability severity prediction, (pre-)patch analysis, automated vulnerability repair, and many others.
We plan to periodically update \CVEfixes, and extend it with mining other open-source projects from
different version control systems and issue tracking systems.
\end{sloppypar}

\section*{Acknowledgments} 
The work presented in this paper has been financially supported by the Research Council of Norway through the secureIT project (RCN contract \#288787).
 
\balance
\printbibliography

\end{document}